# Understanding building blocks of photonic logic gates: Reversible, read-write-erase cycling using photoswitchable beads in micropatterned arrays

Heyou Zhang[1], Pankaj Dharpure[2], Michael Philipp[1],
Paul Mulvaney[3], Mukundan Thelakkat[2,4,5], Jürgen Köhler[1,4,5]

[1]*Spectroscopy of soft Matter, University of Bayreuth, 95440 Bayreuth, Germany*

[2]*Applied Functional Materials, University of Bayreuth, 95440 Bayreuth, Germany*

[3]*ARC Centre of Excellence in Exciton Science, School of Chemistry, University of Melbourne, Parkville, VIC, 3010, Australia*

[4]*Bavarian Polymer Institute, University of Bayreuth, 95440 Bayreuth, Germany*

[5]*Bayreuther Institut für Makromolekülforschung (BIMF), 95440 Bayreuth, Germany*

\* corresponding author: juergen.koehler@uni-bayreuth.de

ORCID:

Mukundan Thelakkat:   0000-0001-8675-1398
Jürgen Köhler:        0000-0002-4214-4008
Paul Mulvaney:        0000-0002-8007-3247
Heyou Zhang:          0000-0001-9513-3874
Michael Philipp:      0009-0007-8717-7317
Pankaj Dharpure:      0000-0001-9975-6921

**Abstract**

Using surface-templated electrophoretic deposition, we have created arrays of polymer beads (photonic units) incorporating photo-switchable DAE molecules, which can be reversibly and individually switched between high and low emission states by direct photo-excitation, without any energy or electron transfer processes within the molecular system. The micropatterned array of these photonic units is spectroscopically characterized in detail and optimized with respect to both signal contrast and cross-talk. The optimum optical parameters including laser intensity, wavelength and duration of irradiation are elucidated and ideal conditions for creating reversible on/off cycles in a micropatterned array are determined. 500 such cycles are demonstrated with no obvious on/off contrast attenuation. The ability to process binary information is demonstrated by selectively writing information to the given photonic unit, reading the resultant emissive signal pattern and finally erasing the information again, which in turn demonstrates the possibility of continuous recording. This basic study paves the way for building complex circuits using spatially well-arranged photonic units.

**Introduction**

Electronic devices are one of the most important inventions of the last century and their impact on everyday life cannot be overestimated. These devices rely on integrated circuits that correspond to a network of interconnected logic gates. A logic gate generates a binary output "1" or "0" depending on the combination of input stimuli and runs on electrons as signal carriers. In contrast, an optical gate may be actuated with the input of light and allows one to exploit not only intensity (i.e. the number of photons per unit time) but also properties such as wavelength or polarisation, thereby enabling opportunities for a massive parallelisation of data transfer. Hence, all optical circuitry remains a long-cherished goal in the fields of high-density data storage and ultrafast communications. Accordingly, the transition from electrical to optical networks has attracted much attention [1–4].

The challenge to develop devices for signal transduction that run on photons rather than on electrons is extremely demanding and requires: materials that feature a high photochemical/photophysical stability, high fatigue resistance, a rapid response, a thermally irreversible bistability, and suitable experimental protocols for immobilizing the photon sources/sinks. All of these bottlenecks have thus far prevented the development of all-optical logic circuits. Photoswitchable molecules are a promising platform for meeting these requirements as they can be interconverted between two bistable, isomeric forms by absorption of light of two different wavelengths [5–7]. This offers the opportunity to optically encode information as the presence or absence of a specific isomer [4,5,8–11]. In the past, several strategies based on photochromism have been proposed to realize all-photonic logic functions for controlling energy and charge transfer efficiencies between photoswitches and dye molecules [12,13]. This has led to the idea of photonic logic gates [14–18]. However, for some of the demonstrated logic functions, the output of one gate corresponds to a change in absorbance at a certain wavelength [15,17], while for others it is necessary to take a spectral scan [15] or even to perform a tedious principal component analysis to determine spectral changes [18]. To date no study has demonstrated an array of gates or controlled spatial immobilization of the photonic units in a solid-state device and, furthermore, none has shown that individual gates can be optically addressed. These are two of the key prerequisites that must be fulfilled in order to utilize these outputs to concatenate two or more logic gates.

Recently, Irie and coworkers introduced a new class of photochromic molecules based on dibenzothienylperfluorocyclopentenes that are intrinsically fluorescent on photocyclization, and which are therefore termed "*turn-on*" photoswitches [19–23]. Accordingly, irradiation with UV light leads to a ring cyclization reaction of the low emitting (LE) open-isomer, transforming it into a highly emitting form (HE) that can be probed with VIS light. Such turn-on photoswitches can be reversibly switched between LE and HE states. However, the light used for probing the emission can also induce the cycloreversion process, resulting in the LE form of the photoswitches. Both processes, i.e., *emission of a photon* and *conversion of the switches to the LE form* are in competition with each other and mutually exclusive. Such "*turn-on*" switches have been used to demonstrate high-security authentication of micrometre-sized optical patterns [24]. Unfortunately, one write-read cycle took several hours to complete. Nevertheless, these results offer micropatterning opportunities with photochromic molecules and are encouraging for the development of all-photonic gates. Ultimate miniaturization envisages the use of a single photochromic molecule for encoding one bit. However, as has been shown recently, individual molecules are generally not suited to encode unequivocal HE/LE states [25,26] due to the unavoidable blinking of single objects [27–30]. Yet, as suggested in [26] a small ensemble of photochromic molecules will be sufficient for an unambiguous discrimination of HE and LE levels.

Here we immobilize an ensemble of photoswitchable 1,2-Bis(2-ethyl-6-phenyl-1-benzothiophene-1,1-dioxide-3-yl)perfluorocyclopentene (abbreviated as DAE) "*turn-on*" photoswitches by incorporating these molecules into polystyrene beads. Then the dye-loaded beads can be actively micropatterned by surface templated electrophoretic deposition (STEPD) within seconds into predefined structures. This allows the fabrication of large-area arrangements of nanoparticles with high precision that can be considered as individual photonic units. In the following, it will be demonstrated that the DAE molecules embedded in polystyrene beads can be switched reversibly hundreds of times between HE and LE states without degradation, and that an array of such DAE-loaded PS beads is fully controllable using light as the external stimulus. The limiting conditions for photoaddressing individual photonic units to achieve i) maximum contrast between HE and LE states that can be associated with the binary outputs "1" and "0", ii) minimum cross-talk between adjacent photonic units, and iii) the feasibility of reversible read-write erase cycles will be systematically explored. By deliberately moving the array of beads relative to the focus of the UV illumination, information can be inscribed sequentially into a group of polymer beads, where each write - read - erase cycle takes just a few seconds. This study of well-structured and spatially immobilized DAE moieties elucidates the interplay of experimental parameters such as illumination intensities, wavelengths, and illumination times for optimum imprinting of information on microstructured matter with optical means and paves the road for the creation of more complex systems based on photonic logic gates.

**Results**

*Photophysics of DAE molecules*

The DAE molecule undergoes reversible isomeric changes upon irradiation with light. Illumination in the UV spectral range (280 nm - 350 nm) induces a photocyclization reaction resulting in a highly emissive (HE) state, whereas illumination in the visible spectral range (420 nm - 520 nm) leads to a photo-cycloreversion reaction, which yields a low-emissive (LE) state, Fig.1a top. For immobilization, the photochromic DAE molecules were incorporated into polystyrene (PS) nanobeads with a diameter of about 310 nm. Thereby, each individual bead accommodates about $10^6$ DAE molecules (see SI). In order to test whether the photophysical properties of the DAE molecules were conserved upon incorporation into the PS nanobeads, the UV/VIS spectra of DAE in toluene solution and embedded in PS in water were compared (for details see SI). As shown in Fig.1b, the emission spectra of DAE in solution and embedded in PS are in close agreement with each other both for the LE and the HE state. This allows clear discrimination between the two emission intensity levels in DAE-doped PS beads.

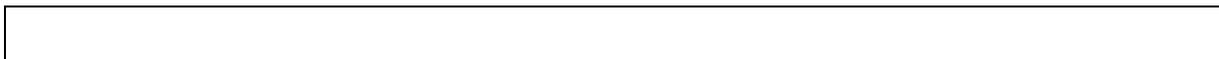

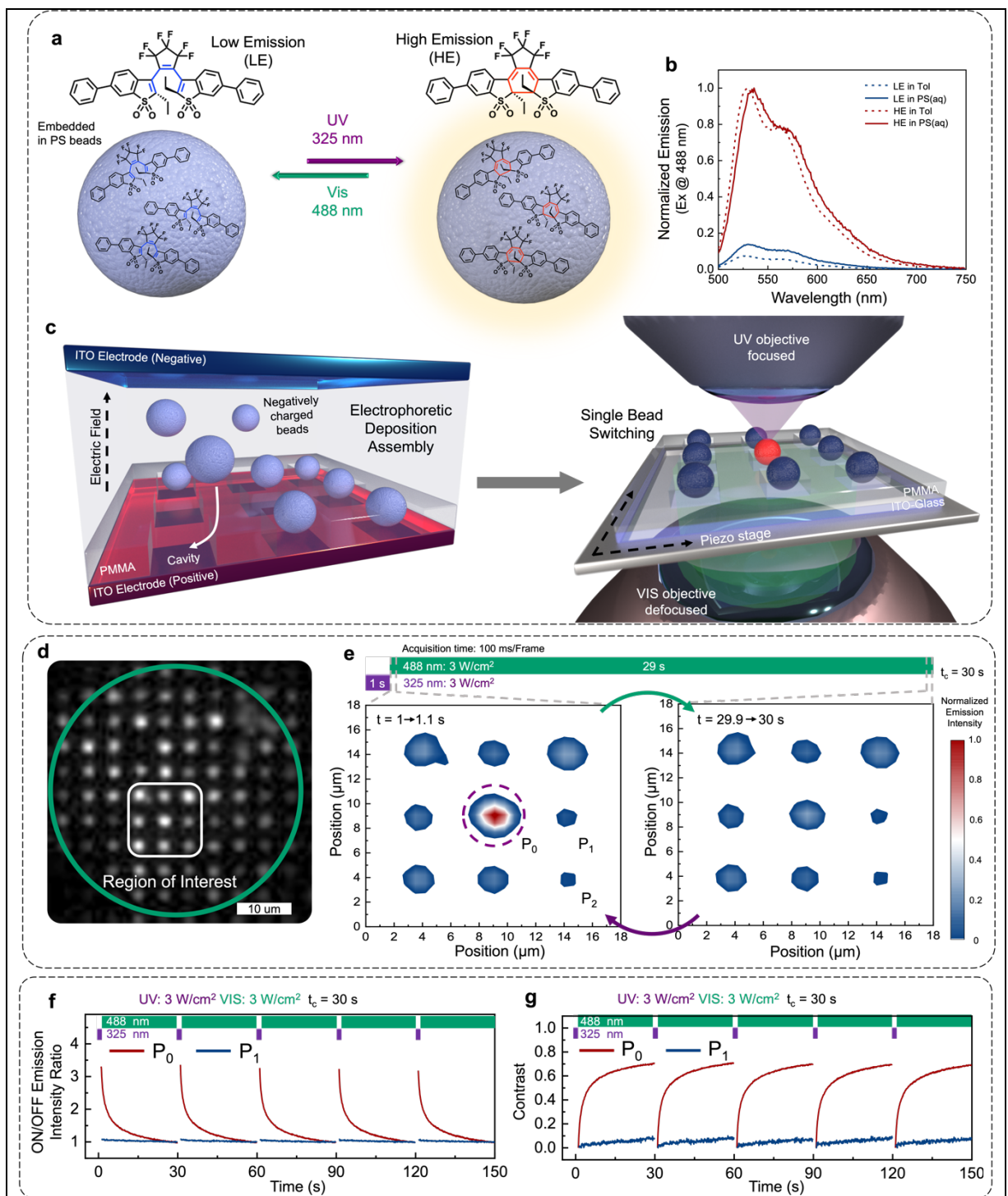

*Fig.1: a) Molecular structure of the DAE molecules in the open, low-emissive (LE) state (left), and in the closed, high-emissive (HE) state (right). Bottom: Sketch of polystyrene (PS) beads (grey spheres) loaded with DAE molecules. The wavelengths used to induce reversible photoswitching are indicated next to the coloured arrows. b) Emission spectra of DAE in toluene (dashed lines) and polystyrene (full lines) in the LE (blue lines) and HE (red lines) states. The excitation wavelength was 488 nm, and the HE spectra have been normalized. The intensity of the LE spectrum is shown relative to the corresponding HE spectrum. c) Left: Schematic sketch of a sample cell for STEPD. The cell consists of a PMMA substrate (PMMA thickness 50 nm) with predefined cavities (500 nm x 500 nm) prepared with electron beam lithography and is placed between positively and negatively charged ITO electrodes. Right: Sketch of the optical experiment exposing the beads to UV (from the top) and VIS (from the bottom) radiation via two microscope objectives. While the VIS light illuminates a*

> larger area of the sample (70 µm in diameter), the UV light is focussed to a spot size of roughly 500 nm in diameter and can be superimposed selectively on individual nanobeads by moving the whole sample with a piezo stage. d) Example of an array of PS nanobeads arranged on a 5 µm × 5 µm grid with the STEPD method. The green line indicates the area that is illuminated by the visible laser at 488 nm, and the white rectangle corresponds to the region of interest (ROI) that is studied further. e) Fluorescence images of the ROI for an optical cycle of 1s illumination at 325 nm (3 W/cm$^2$) and 29s of illumination at 488 nm (3W/cm$^2$), as indicated by the bars at the top. The images have been acquired at the beginning (left) and the end (right) of the VIS illumination period for bin times of 100ms. In the left part the purple dashed line indicates the area that is irradiated with UV light. The fluorescence intensity has been normalized to the HE level and is colour coded, see bar at the right. f) Response of the normalized emission intensity (left axis) as a function of time for the beads $P_0$ (red line) and $P_1$ (grey line) for five consecutive illumination cycles. g) Contrast C(t) for the beads $P_0$ (red line) and $P_1$ (grey line) for five consecutive illumination cycles.

*Photoaddressing micropatterned arrays*

Next the dye-loaded beads have been assembled into a two-dimensional lattice combining electrophoretic deposition (EPD) and electron beam lithography (EBL), referred to as surface-templated electrophoretic deposition (STEPD). With STEPD electrically charged nanoparticles can be positioned under the influence of an external electric field onto a predefined template that has been prepared with EBL, Fig.1c. This allows the creation of large-area assemblies of nanoparticles with high precision. A detailed description of the method can be found in [31]. We used these techniques to position individual DAE-loaded PS beads into predefined cavities that were crafted by EBL to form a grid with 5 µm spacing, Fig.1d. Based on dynamic light scattering and zeta-potential measurements, the DAE-loaded PS beads feature a hydrodynamic size of 334 nm ± 20 nm and a zeta-potential of −33.9 mV ± 0.5 mV (see SI). Hence, the cavity size was fabricated to be 500 nm × 500 nm, and a positive potential (+4 V with respect to counter ITO) was applied during the STEPD assembly. The sample was attached to a piezo stage for precise movements of the array and the cell was mounted in a home-built, dual-light, microscope setup for switching and exciting single beads. The green circle in Fig.1d refers to the area of about 70 µm in diameter that is illuminated with the VIS light, and the white square refers to a region of interest (ROI) that accommodates 3 × 3 polymer beads. Fluorescence images from this area are shown in Fig.1e on an expanded scale. These were obtained by focussing the UV light for 1 s onto the PS bead $P_0$ in the centre (dashed purple line) and subsequently probing the fluorescence from all nine PS beads by excitation with VIS light for 29 s (i.e. the total duration of one cycle is 30 s). The two images in Fig.1e have been accumulated during the first 100 ms (left) and the last 100 ms (right) of exposure to VIS light. Only for the bead $P_0$, whose chromophores have been initialized in the HE state, is a strong fluorescence signal observed upon illumination with VIS light, Fig.1e left, whereas the signal from the other beads remains at the low level. Importantly, as the VIS illumination proceeds, the emission intensity registered from the bead $P_0$ decays to the LE level on a timescale of several seconds, Fig.1e right. This is compared more quantitatively in Fig.1f for $P_0$ and one of its nearest neighbouring beads, $P_1$. The figure shows the normalized emission intensity of these beads for five consecutive UV-VIS illumination cycles of 30s duration each. The high degree of reproducibility of the five cycles illustrates the reversibility of the switching process. Taking another point of view, this can be interpreted as a repetition of purely optical *write-read-erase* cycles. The UV light writes a piece of information into the nanobead that can be read out by fluorescence upon VIS excitation. Prolonged VIS excitation erases this information and clears the bead for a new writing cycle. In order to quantify this optical cycle more precisely, we define the contrast as a function of time as

$$C(t) = \frac{I_{max} - I(t)}{I_{max}} = 1 - \frac{I(t)}{I_{max}}$$

where $I_{max}$ refers to the emission intensity that is obtained at the beginning of the 488 nm illumination period, and $I(t)$ refers to the intensity at time *t* during the 488 nm illumination period. Since the VIS radiation also converts the DAE molecules to the LE state, $I(t)$ decreases as a function of time, and the maximum contrast between the HE and LE levels is obtained for long 488 nm illumination times. In other words, erasing the information from the previous writing cycle is more complete, the longer the bead is exposed to VIS light. For the beads $P_0$ (black line) and $P_1$ (grey line) the temporal development of the contrast is compared in Fig.1g. For $P_0$ it shows a steep rise from 0 to 0.5 within a few seconds at the beginning of the 488 nm illumination time and saturates at longer times at about 0.7. For $P_1$ this parameter shows only a small change as a function of time from 0 to about 0.1 after 30 s.

*Elucidating parameters for optimum performance*
Obviously, the contrast that can be achieved for a particular bead is a function of the illumination intensities of the two conversion beams and the duration of the cycle time. For optical writing/reading/erasing information it is therefore important to find the best combination of these parameters to achieve the largest discriminatory power between the signals. Yet, before this is done, we first address the issue of crosstalk between neighbouring beads. In Fig.2a, the ROI of Fig.1d is shown on an enlarged scale specifying the spatial separations between nearest-neighbour beads (5 µm) and next-nearest-neighbour beads (7.1 µm). As before, only the PS bead in the centre, $P_0$, is irradiated with UV light, converting the incorporated DAE molecules to the closed (HE) conformation.

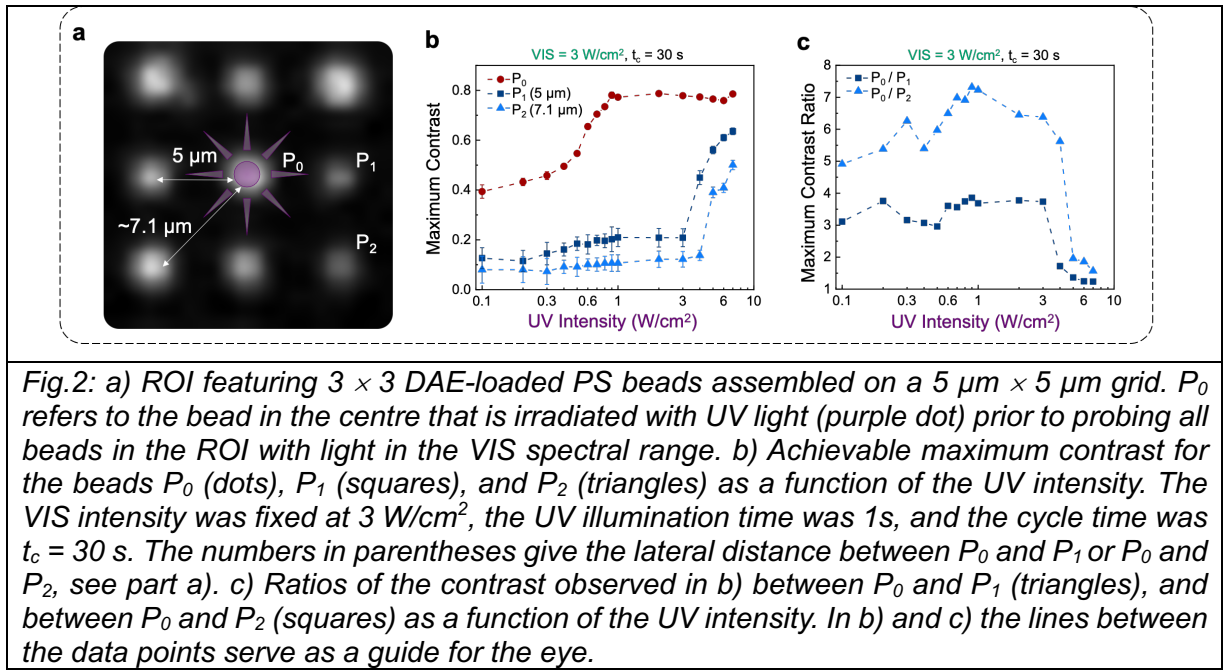

*Fig.2: a) ROI featuring 3 × 3 DAE-loaded PS beads assembled on a 5 µm × 5 µm grid. $P_0$ refers to the bead in the centre that is irradiated with UV light (purple dot) prior to probing all beads in the ROI with light in the VIS spectral range. b) Achievable maximum contrast for the beads $P_0$ (dots), $P_1$ (squares), and $P_2$ (triangles) as a function of the UV intensity. The VIS intensity was fixed at 3 W/cm², the UV illumination time was 1s, and the cycle time was $t_c$ = 30 s. The numbers in parentheses give the lateral distance between $P_0$ and $P_1$ or $P_0$ and $P_2$, see part a). c) Ratios of the contrast observed in b) between $P_0$ and $P_1$ (triangles), and between $P_0$ and $P_2$ (squares) as a function of the UV intensity. In b) and c) the lines between the data points serve as a guide for the eye.*

For a fixed VIS excitation intensity of 3 W/cm², Fig.2b shows the maximum contrast as a function of the UV excitation intensity for the bead $P_0$, one of its nearest neighbours, $P_1$, and one of its next-nearest neighbours, $P_2$. For the beads $P_1$ and $P_2$ whose chromophores have not been converted to the HE state, the contrast remains almost constant at C = 0.1 up to an excitation intensity of 3 W/cm². For excitation intensities above this value the contrast for these beads rises to about 0.6. This rise is presumably caused by scattering of the UV radiation toward the beads that are not within the UV focus, resulting in the conversion of some of the DAE molecules to the HE state. This is in contrast to the findings for $P_0$, whose molecules were initialized to the HE state. For this bead the contrast already begins to increase at 0.4 for low UV excitation intensities and shows a significant rise to 0.75 around 1 W/cm². Dividing the observed contrasts of $P_0$ and $P_1$ ($P_2$) yields the contrast ratios shown in Fig.2c. This reveals ratios above 5 ($P_0/P_1$) and above 3 ($P_0/P_2$) for UV excitation intensities below 3 W/cm², which then both drop to about 1 for higher UV intensities. Hence, the system exhibits very strong discriminatory power between the HE and LE levels, and at the same time low crosstalk for UV intensities between 1 - 3 W/cm².

Finally, in order to identify the optimum operational parameters and also to give an over view of photonic unit switching performance, all combinations of excitation intensities ranging from 1 - 6 W/cm$^2$ for the VIS and 2 - 7 W/cm$^2$ for the UV (both in increments of 1 W/cm$^2$) have been tested for cycle times $t_c$ of 5 s, 10 s, 20 s, 30 s, 40 s, and 50 s. This yields in total six 6 × 6 intensity matrices for the maximum contrast and these are shown in Fig.3a. For nearly all applied intensities the contrast rises monotonically as a function of the cycle time because the longer the VIS illumination time, the greater the number of DAE molecules that are converted back to the LE state. For VIS intensities between 3 - 6 W/cm$^2$ the observable contrast does not depend much on the VIS intensity, and the largest growth of the contrast is observed during the first 30 s (from about 0.2 to about 0.8), followed by a further minor increase to about 0.9 after 50 s. For the shortest cycle time of 5 s, it turns out that the maximum achievable contrast is generally poor. For all combinations of excitation intensities, it does not exceed a value of 0.4, and even values above 0.3 can only be achieved at the highest excitation intensities, which are problematic with respect to crosstalk, see above. This reflects the fact that the VIS illumination time of 4 s is not long enough to convert a sufficient fraction of DAE molecules back to the LE state. Larger maximum contrasts can be obtained for higher intensities and longer cycle times. For example, for the highest UV and VIS intensities and 50s cycle time we find C = 0.93.

Given that the UV intensity should not exceed 3 W/cm$^2$ to diminish crosstalk, and that an increase in the cycle time from 30 s to 50 s yields only a moderate increase in the maximum contrast, we have used a cycle time of $t_c$ = 30 s and excitation intensities of 3W/cm$^2$ for both wavelengths to conduct further experiments. This choice is a good compromise between high contrast, suppression of crosstalk, and the duration of the experiment. For this setting of the parameters Fig.3b shows the contrasts C(t) for 500 consecutive illumination cycles. For all individual cycles the maximum contrast exceeds 0.7 and does not show any degradation during the more than four hours of total illumination time, confirming the high photostability of these novel, chemical switching systems.

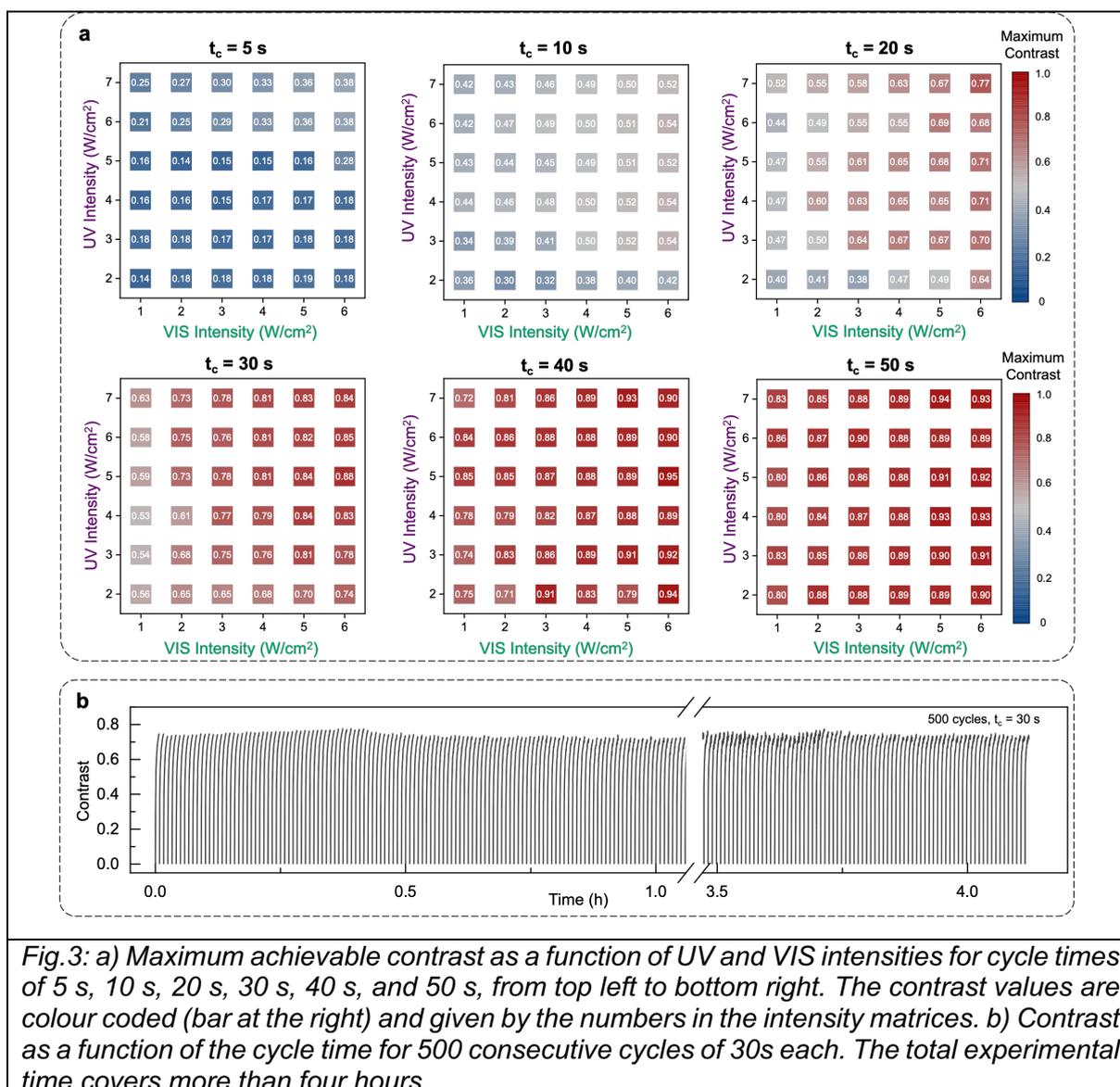

*Fig.3: a) Maximum achievable contrast as a function of UV and VIS intensities for cycle times of 5 s, 10 s, 20 s, 30 s, 40 s, and 50 s, from top left to bottom right. The contrast values are colour coded (bar at the right) and given by the numbers in the intensity matrices. b) Contrast as a function of the cycle time for 500 consecutive cycles of 30s each. The total experimental time covers more than four hours.*

The results shown so far demonstrate that the DAE-loaded PS beads within the array can be addressed selectively for photochromic switching. Hence, information can be encoded into the pattern by moving the beads selectively, one by one, into the UV focus, Fig.1c right.

*Reversible write - read - erase cycling*

With the characterization shown above at hand we now have the ingredients to demonstrate purely optical, write-read-erase cycles on micro-patterned surfaces. The standard procedure applied is as follows (see Fig.4a,b): 1) Initialization: Because the DAE molecules in the LE state give rise to a low fluorescence signal, the sample area of interest is illuminated for 2 s with VIS light at 3 W/cm$^2$ to determine the LE background. 2) Writing: With the aid of a piezo-stage a PS bead is moved into the focus of the UV light and irradiated for 1 s (3 W/cm$^2$). This converts the DAE molecules of this particular bead into the HE state. Then the next bead is selected and brought for 1 s into the UV focus by moving the piezostage, (see Fig.4b centre). This is repeated for as many PS beads as are required to encode specific information into the microarray of PS beads (see movie in the SI). At the end of this procedure the array contains a selected group of PS beads that contain DAE molecules "sleeping" in the HE state. 3) Read-Out: The whole sample area is illuminated for 1 s with VIS light (3 W/cm$^2$). This results in a strong fluorescence from those PS beads whose DAE molecules had been converted into the HE state (Fig.4b right). 4) Erasure: The sample is illuminated for 18 s with VIS light at an intensity of 9 W/cm$^2$. The increased VIS intensity is applied in order to accelerate the recovery of the DAE molecules from the HE back to the LE state. It is important to note that the VIS intensity applied during Initialization (step 1) and Read-Out (step 3) have to be identical in order to correctly take the background into account. As a simple proof of the efficacy of this optical encoding system, the letters of the alphabet have been written consecutively onto the ***same*** 5 × 5 PS array of beads in the sample, see Fig.4c.

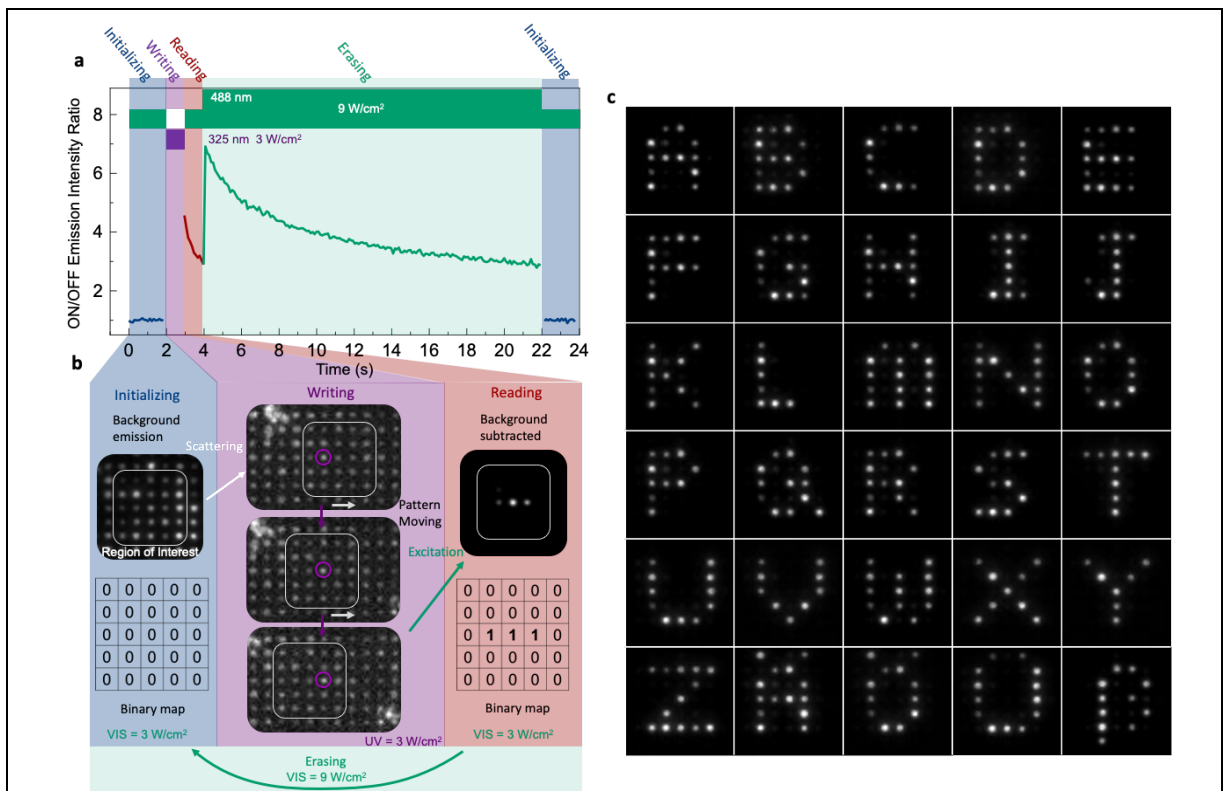

*Fig.4: a) Timing sequence for initialization (blue, VIS, 2 s, 3 W/cm$^2$), writing (purple, UV, 1 s, 3 W/cm$^2$), read-out (red, VIS, 1 s, 3 W/cm$^2$), and erasure (green, VIS, 18 s, 9 W/cm$^2$). b) Sample area (ROI) of 5 × 5 DAE loaded PS beads illustrating the initialization (blue, left), writing (purple, centre), read-out (red right), and erasure (green bottom) processes. For writing, the array of PS beads can be moved relative to the focal spot of the UV light, which allows the conversion of the DAE molecules in several beads, one at a time for 1 s, to the*

*HE level. For read-out the sample is probed with VIS light and the fluorescence signal is accumulated during the first 100 ms of the read-out period. Only those beads whose chromophores have been converted to the HE level show a fluorescence signal above background. The corresponding binary output maps are shown as well. c) For illustration the letters of the alphabet have been written consecutively, one after the other, onto the same area of the sample.*

## Discussion

Over the last few decades much attention has been given to the demonstration of the function of molecular logic circuits [1,4]. However, very often, at least some of the transitions between the molecular states were induced using diffusion-limited processes such as enzymatic reactions, pH changes, and/or temperature changes that slow down the speed of operation [32,33]. This has inspired many researchers to investigate the potential of all-optical data processing, i.e. using light as an external stimulus [2,34]. Although the basic functionality of an optical transistor was demonstrated in several benchmark experiments, the systems used featured either low gain [35,36], were operated at cryogenic temperatures [36], relied on non-linear, light-matter interactions needing strong electric fields [37], i.e. very high light intensities, or they consisted of self-assembled DNA nanostructures [38].

Beyond proof-of-principle experiments, the realization of photonic logic gates, i.e. circuits that run on photons rather than electrons, requires building blocks that allow reversible generation of binary outputs with high discriminatory power. In [39] reversible writing, reading, and erasing of information by optical means was demonstrated. Yet, each individual step took several minutes and the sample was illuminated through a photomask. Alternatively, in [40] confocal illumination techniques were used to write different images within seconds repetitively onto a surface that consisted of a photoswitchable, inorganic-organic hybrid system. However, since logic circuits require concatenation and this in turn requires light from one source to control and manipulate light from another source, precise positioning of the switchable units with respect to each other is crucial. This necessitates actively patterned surfaces for control and spacing of the spatial arrangement of the photoswitchable units. As a first step towards this goal, we have demonstrated that micropatterned arrays of polymer beads loaded with photochromic molecules can be prepared and addressed selectively to inscribe information. We have elucidated the interplay of external stimuli, such as the illumination intensities and illumination times for writing, reading, and erasing of information for achieving the optimum contrast and minimum crosstalk. The functionality of this approach is exemplified by consecutively writing, reading and erasing the letters of the alphabet onto the same 25 polymer beads arranged in a 5 μm × 5 μm matrix, which took only a few seconds per step. Taking advantage of the large variety of derivatives of photochromic DAE molecules, this can be exploited to create variations of the emission wavelengths for particular photonic units. This provides spectral selectivity as a further parameter. Together with modern patterning techniques, this opens a route for the concatenation of several photonic units for accomplishing interconnected photonic logic gates. In summary, these results demonstrate that photochromism can be used to undertake logic functions using photons as signal carriers.

**Acknowledgement**

We thank Yue Dong for assistance with the electron microscopy and Max Gießübel for measuring the absorption spectra required to calculate the DAE content of the polymer beads. JK and MT thankfully acknowledge financial support by the Deutsche Forschungsgemeinschaft (Ko 1359/30-1, TH 807/11-1, GRK 2818) and the Bavarian state ministry for arts and science within the initiative "Solar Technologies go Hybrid". PM thanks the ARC for support through CE170100026.

Supporting Information for

# Understanding building blocks of photonic logic gates: Reversible, read-write-erase cycling using photoswitchable beads in micropatterned arrays

Heyou Zhang[1], Pankaj Dharpure[2], Michael Philipp[1], ,
Paul Mulvaney[3], Mukundan Thelakkat[2,4,5], Jürgen Köhler[1,4.5]

[1]*Spectroscopy of soft Matter, University of Bayreuth, 95440 Bayreuth, Germany*

[2]*Applied Functional Materials, University of Bayreuth, 95440 Bayreuth, Germany*

[3]*ARC Centre of Excellence in Exciton Science, School of Chemistry,
 University of Melbourne, Parkville, VIC, 3010, Australia*

[4]*Bavarian Polymer Institute, University of Bayreuth, 95440 Bayreuth, Germany*

[5]*Bayreuther Institut für Makromolekülforschung (BIMF), 95440 Bayreuth, Germany*

Table of Contents

## 1. Synthesis of the photoswitches

The phenyl derivative, 1,2-bis(2-ethyl-6-phenyl-1-benzothiophene-1,1-dioxide-3-yl)perfluorocyclopentene (DAE) was synthesized by adapting published procedures. The respective non-oxidized forms were synthesized by elimination reactions of perfluorocycloalkenes with organolithium compounds as described in [1]. To obtain DAE with ethyl substituents, oxidation using $H_2O_2$ was adopted [2]. To obtain the DAE photoswitch, the oxidized compound, 1,2-bis(2-ethyl-1-benzothiophene-1,1-dioxide-3-yl)perfluorocyclopentene was first iodinated as shown below, followed by a Suzuki-Miyaura coupling reaction using the 6,6´-diiodo derivative and phenylboronic acid.

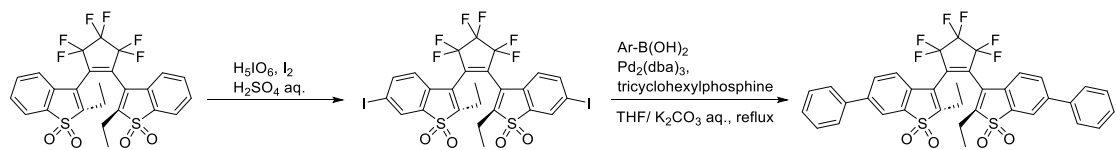

Synthesis: Dissolve 1,2-Bis(2-ethyl-6-iodo-1-benzothiophene-1,1-dioxide-3yl)perfluorocyclopentene (0.5 g, 0.616 mmol) and phenylboronic acid (0.188 g, 1.54 mmol) in 33 mL THF, add 10 mL aq. saturated solution of $K_2CO_3$., then add $Pd_2(dba)_3$ (0.1 g, 0.11 mmol), then 0.33 mL tricyclohexylphosphine (18% in toluene solution). Purge the reaction mixture with argon for 15 min and reflux for 12 h at 75 °C. Upon completion of reaction, filter the reaction mixture through celite. Extract three times with chloroform, dry the organic phase over $Na_2SO_4$, then remove the solvent with an evaporator. Purify using column chromatography (n-Hex/EA 9:1), followed by freeze-drying with dioxane and vacuum drying for 24 h at 80°C. Yield: 49% dark yellow solid.

$^1$H NMR (300 MHz, CDCl$_3$) δ: 7.97 (d, J = 1.7 Hz, 1.2 H), 7.90 (d, J = 1.7 Hz, 0.8 H), 7.82 (dd, J = 8.0, 1.7 Hz, 1.2 H), 7.65 – 7.56 (m, 3.4 H), 7.54 – 7.39 (m, 7.9 H), 7.27 (d, J = 8.4 Hz, 1.2 H), 7.25 – 7.20 (m, 0.9 H), 2.73 – 2.52 (m, 2.8 H), 2.44 (m, 1.2 H), 1.43 (t, J = 7.6 Hz, 2.4 H), 1.10 (t, J = 7.6 Hz, 3.6 H). $^{13}$C NMR (75 MHz, CDCl$_3$) δ: 148.6, 148.2, 144.6, 144.5, 138.2, 138.1, 136.5, 136.5, 132.3, 131.9, 129.4, 129.3, 129.2, 129.1, 128.2, 128.1, 127.2, 127.2, 123.5, 123.4, 123.2, 123.0, 121.2, 19.4, 19.3, 12.1, 11.8; MALDI-TOF MS (EI) m/z (M+) calculated for $C_{37}H_{26}F_6O_4S_2$: 712, found 712.

*2. Optical spectroscopy*

The UV/VIS spectra of DAE in toluene as well as of an aqueous solution of DAE embedded in PS were recorded with commercial spectrometers (PerkinElmer Lambda 750 UV/VIS Spectrometer; Cary Eclipse Fluorescence Spectrometer).

The micropatterned samples were mounted in a home-built optical microscope on a piezo stage (Nano-Drive™, Nano-LP100(Stage), ND3-USB203 (Controller), Mad City Lab Inc.). The light sources were either a helium-cadmium laser (IK3201R-F, Kimmon) for 325 nm, or a laser diode (Cyan Laser Head, Spectra Physics) for 488 nm. The UV light was cleaned up by a bandpass filter (320 nm, bandwidth 40 nm, AHF), converted to circular polarization (quarter waveplate, Thorlabs) and focused from the top by an objective (LUCPlanFL N 60x/NA 0.70, Olympus) to a spot size of about 500 nm in diameter. The VIS light was also converted to circular polarization (quarter waveplate, Thorlabs), reflected by a dichroic beam splitter (reflection 350-458 nm > 90%, transmission 464-900 nm > 93%, AHF), and weakly focused by another objective (LD Plan-NEOFLUAR 20x/NA 0.4 Korr, Zeiss) from the bottom of the microscope to a spot size of 70 µm in diameter. For both light sources the excitation intensity was monitored with a power meter (325 nm: FieldMax II – TO (ROHS), Coherent; 488 nm: PM100A, Thorlabs.) and adjusted with variable OD filters (OD 0.04-3 (reflective), Edmund Optic.). The emission from the sample was collected via the bottom branch of the microscope, transmitted through the dichroic mirror, filtered by a bandpass (578nm, bandwidth 105 nm, AHF) and detected with an electron multiplying CCD camera (iXon, DV887ECS-UVB, Andor). In addition, the sample could be illuminated with an LED lamp (H-R3, XP-E3 LED, Ultrafire.) in the red spectral range for taking images of the array from the scattered light. In order to prevent detection of light directly from the LED, the illumination path was inclined by 30° with respect to the sample plane. The signal was detected with a programmable real-time I/O system (ADWin Gold II, Jäger Messtechnik), that was also used for controlling electromechanical shutters (Uniblitz, temporal accuracy 5 ms). These ensured the sequential illumination of the sample. The setup for investigating the micropatterned samples is shown schematically in Fig.S1.

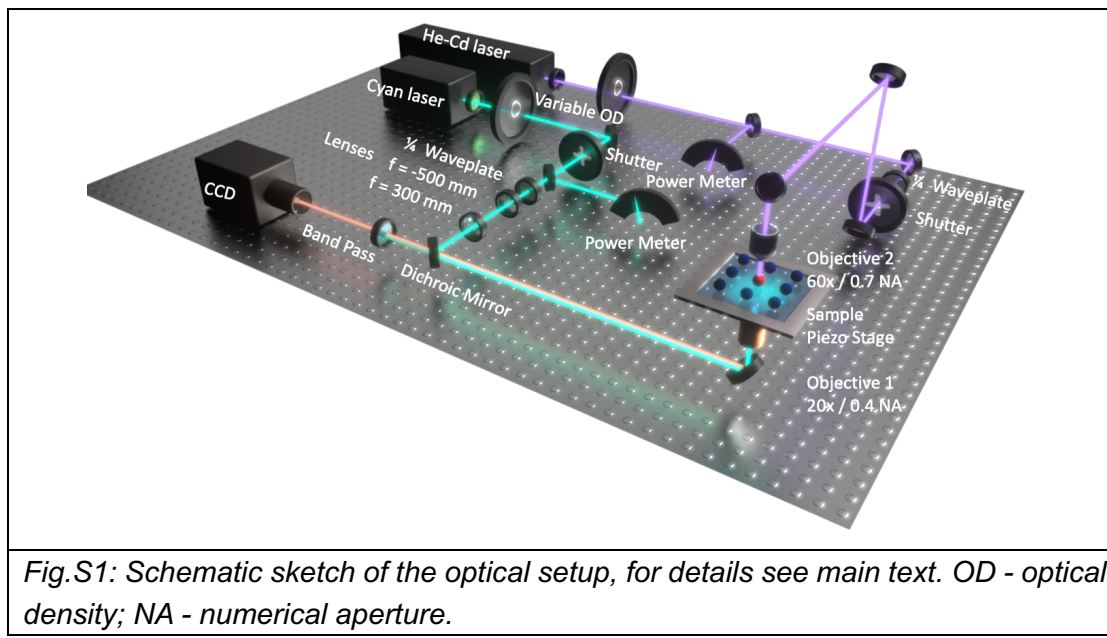

Fig.S1: Schematic sketch of the optical setup, for details see main text. OD - optical density; NA - numerical aperture.

### 3. Incorporation of DAE into polymer beads as photonic units

To prepare a stock solution, 2 mg of DAE powder were dissolved in 2 ml of tetrahydrofuran (THF). Polystyrene beads of about 300 nm in diameter were purchased (Bangs Laboratories Inc, PS02009, 10.05%, 100mg/ml) as a colloidal solution in water and diluted 100 times using Milli-Q water. Then 1 ml from this solution was mixed with 1 ml of Milli-Q water, and 1 ml of THF, by mild magnetic stirring for 5 minutes. Subsequently, 0.2 ml from the DAE stock solution were further diluted with THF to 1 ml and added to the mixture under rigorous stirring for 5 minutes. The mixture was then kept for approximately 12 hours in the dark at room temperature. In order to incorporate the DAE molecules into the PS beads, the mixture was loaded onto a Schlenk line, and about 2 ml of the solvent was evaporated within 20 minutes under low vacuum. The remaining material was centrifuged (6100 rpm / 2500 rcf (relative centrifugal force), Eppendorf® MiniSpin®) for 30 mins, and after removing the supernatant, the precipitate was redispersed in Milli-Q water. The centrifugation was repeated three times, and the resultant colloidal solution was dispersed in 3ml Milli-Q water. Finally, the beads were characterized by transmission-electron microscopy, dynamic light scattering, and measuring the zeta potential, (see below).

### *4. Structural characterization of the DAE-loaded polystyrene beads*

Transmission electron microscopy (TEM) was performed on a Zeiss Libra 120 and Libra 200 (Zeiss, Germany), using an acceleration voltage of 120 kV. The samples were prepared by drop casting 2 µl of the colloidal solution (0.2 µg/ml) of DAE-loaded PS beads on the carbon foil coated side of the copper grids, and air-dried, Fig.S2.

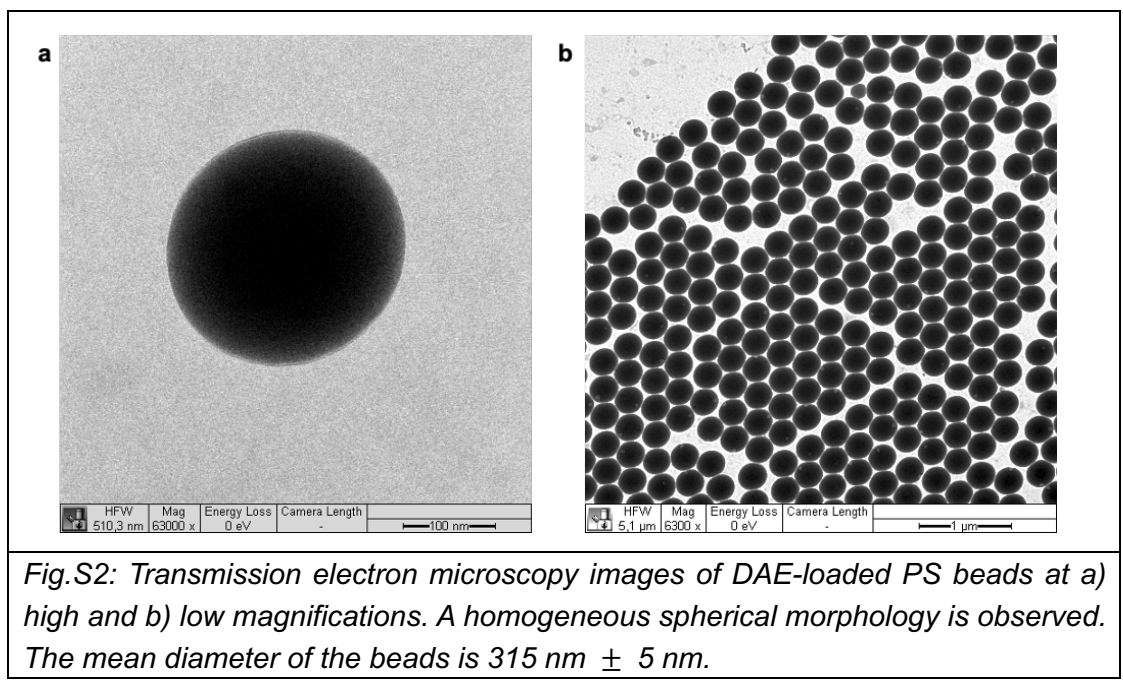

*Fig.S2: Transmission electron microscopy images of DAE-loaded PS beads at a) high and b) low magnifications. A homogeneous spherical morphology is observed. The mean diameter of the beads is 315 nm $\pm$ 5 nm.*

Dynamic light scattering (DLS) and zeta potential measurements of the STEPD-ready solution (DAE-loaded PS beads (0.15 µg/ml) with NaCl (0.2 mmol/l) in water) were conducted using a Zetasizer (Malvern Instruments, Inc., Malvern, U.K.). The size of the DAE-loaded PS beads is given as intensity-average diameter values assuming a refractive index of 1.59, Fig.S3.

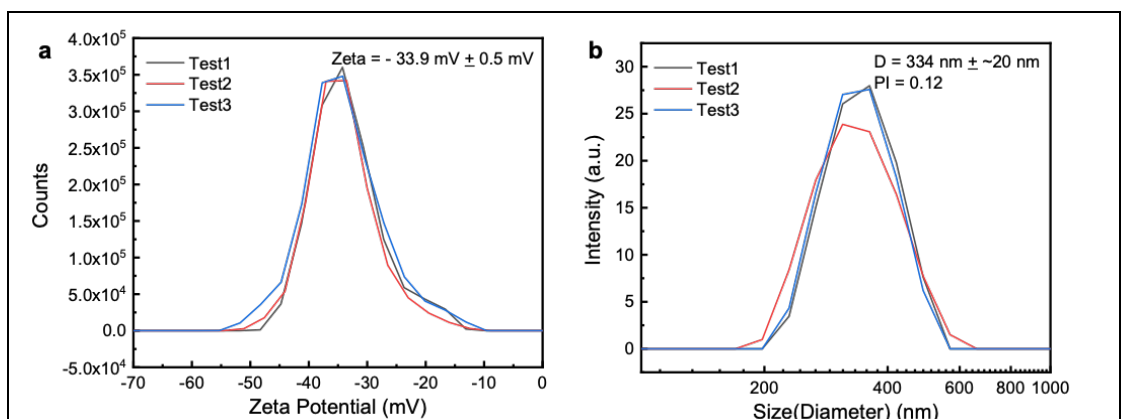

*Fig.S3: a) Zeta potential and b) dynamic light scattering characterization of DAE-loaded PS beads as prepared for STEPD assembly (0.2 mmol/l NaCl, pH = 7, room temperature). The zeta potential of the beads is – 33.9 mV $\pm$ 0.5 mV, and the hydrodynamic diameter of the beads is 335 nm $\pm$ 20 nm in good agreement with TEM.*

## 5. Spectroscopic characterization of the DAE-loaded PS beads

Illumination of DAE in the UV spectral range (280 nm - 350 nm) induces a photocyclization reaction resulting in a highly emissive (HE) state, Fig.S4a top right, whereas illumination in the visible spectral range (420 nm - 520 nm) leads to a photo-cycloreversion reaction, which yields a low-emissive (LE) state, Fig.S4a top left. In order to test whether the photophysical properties of the DAE molecules were conserved upon incorporation into the PS nanobeads, the UV/VIS spectra of DAE in toluene solution and embedded in PS in water are compared in Fig.S4b,c. In toluene, DAE shows the well-known absorption bands around 300 nm - 400 nm for the open, low-emissive (LE) state, and around 400 nm - 520 nm for the closed, high-emissive (HE) state. For DAE in PS these characteristic features are reproduced on top of a strong background due to light scattering that stems from the PS, and which decreases monotonically from short to long wavelengths, Fig.S4b. As shown in Fig.S4c, the emission spectra of DAE in solution and embedded in PS are in close agreement with each other both for the LE and the HE states.

Next, the modulation of the emission from the DAE molecules within the PS nanobeads was investigated. As an example, Fig.S4d shows two fluorescence images of an area of 90 μm × 90 μm that contains four DAE-loaded polymer beads that were spread randomly on the substrate. These were illuminated first with UV light for 1s and then with VIS light for 19s, as indicated by the coloured bars at the top of Fig.S4d. The 325 nm radiation was focussed to a diameter of 500 nm and positioned onto exactly one of the polymer beads (Fig.S4d left, dashed purple circle, dimension not to scale), referred to as $P_0$, whereas the radiation at 488 nm was only weakly focussed to a spot size of about 70 μm in diameter that covered all four polymer beads (Fig.S4d, green circle). The three images shown in Fig.S4d were recorded at different moments during the illumination cycle, namely during the first 1s of UV illumination, Fig.S4d left, and then during the first 100 ms of the VIS illumination, Fig.S4d centre, and finally during the last 100 ms of the 19s VIS illumination period, Fig.S4d right, respectively. One illumination cycle covers one exposure with radiation at 325 nm, followed by one exposure with radiation at 488 nm.

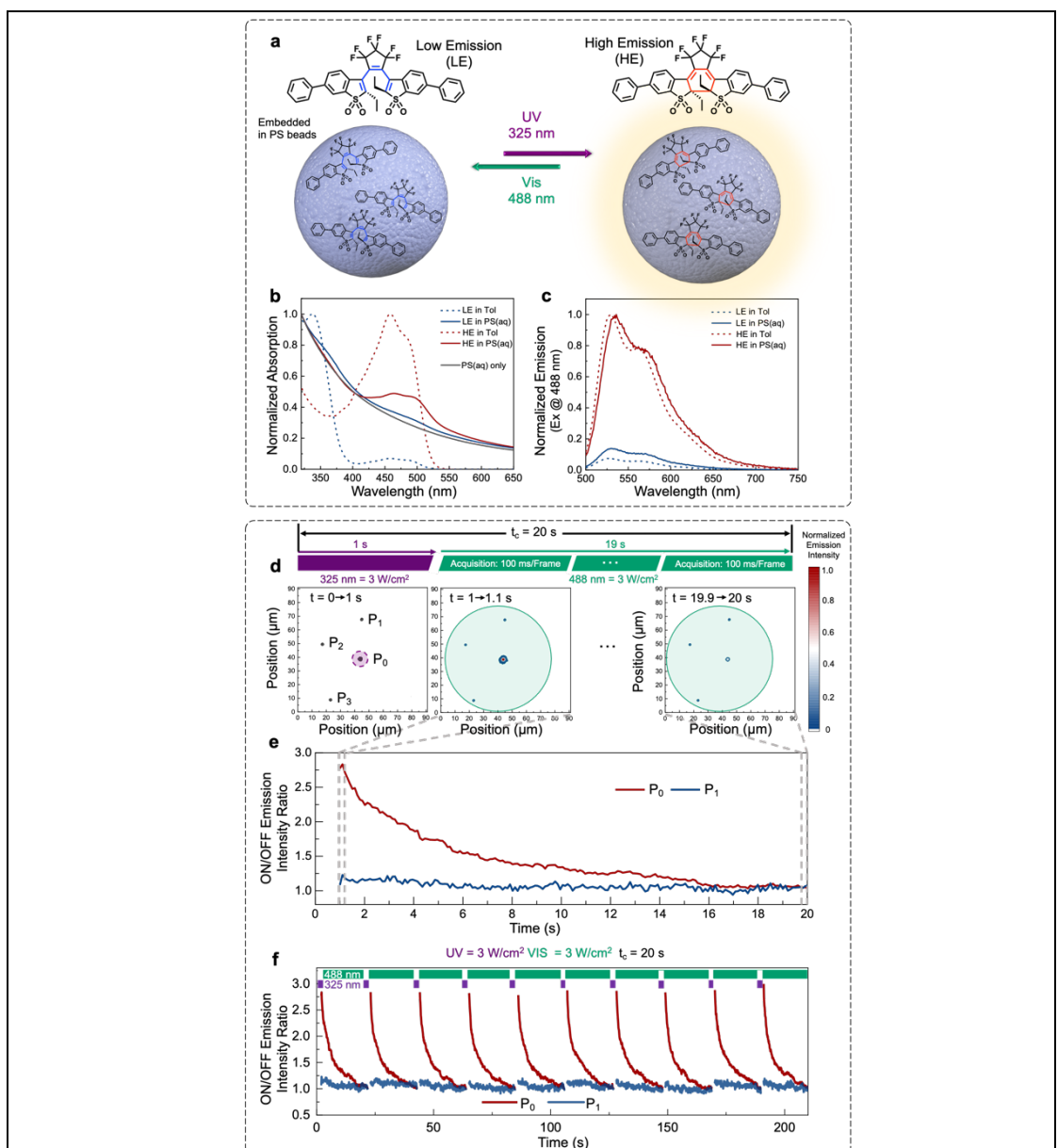

*Fig.S4: a) Top: Molecular structure of the DAE molecules in the open, low-emissive (LE) state (left), and in the closed, high-emissive (HE) state (right). Bottom: Sketch of polystyrene (PS) beads (grey spheres) loaded with DAE molecules. The wavelengths used for inducing the reversible photoswitching are indicated next to the coloured arrows. b) Absorption and c) emission spectra of DAE in toluene (dashed lines) and in PS beads (full lines) in the LE (blue lines) and HE (red lines) states. For comparison the black line shows the absorption of PS beads alone. For better comparison all spectra have been normalized. C) Emission spectra of DAE in toluene (dashed lines) and polystyrene (full lines) in the LE (blue lines) and HE (red lines) states. The excitation wavelength was 488 nm, and the HE spectra have been normalized. The intensity of the LE spectra is shown relative to the corresponding HE spectrum. D) Fluorescence image of four PS nanobeads $P_0 – P_3$ randomly drop cast on a glass substrate. In the left panel the dashed purple circle indicates the focus of the UV beam (diameter 500 nm, dimension not to scale) that was centred on $P_0$, and in both panels the green circle (diameter 70 μm) indicates the focal area*

*of the beam at 488 nm that covers all four beads. One illumination cycle of 20s covers exposure to radiation at 325 nm (3W/cm$^2$) for 1s followed by exposure to radiation at 488 nm (3W/cm$^2$) for 19s, as indicated by the coloured bars at the top. The image in the centre was acquired during the first 100 ms of VIS illumination, and the image on the right-hand side was acquired during the last 100 ms of VIS illumination. For both images the normalized intensity of the emission is given by the colour code (bar at the right). E) Emission intensity of beads $P_0$ (red line) and $P_1$ (blue line) as a function of VIS illumination time for one cycle of sequential illumination. The intensities are normalized to the emission intensity in the LE state. All data points have been acquired for time intervals of 0.1s. f) Sequence of 10 illumination cycles for the same experimental conditions as in e).*

Irradiating $P_0$ for 1s at 325 nm initializes the DAE molecules in the HE state and subsequent excitation at 488 nm results in strong fluorescence from this bead. Because only the DAE molecules in $P_0$ are converted to the HE state prior to probing the emission, only this bead exhibits strong fluorescence, while the DAE molecules in the other three beads still reside in the LE state, and hence exhibit only weak emission, Fig.S4d centre. Nevertheless, after 18.9s of illumination at 488 nm the emission from $P_0$ has also decreased to a lower level, Fig.S4d right. This is shown in more detail in Fig.S4e that features the emission intensity normalized to the LE level, as a function of the VIS illumination time for the beads $P_0$ and $P_1$. For the bead $P_1$ whose DAE molecules have not been initialized in the HE state, the relative emission intensity does not change as a function of the 488 nm illumination time and the PL remains at its LE intensity level over the full cycle time of 20s. In contrast, for the $P_0$ bead, in which the dye molecules have been converted to the HE state, the relative emission intensity is enhanced by a factor of about 2.8 with respect to its emission in the LE state upon starting the VIS illumination, and this PL then decays to the LE level intensity on a timescale of some seconds. This decrease is easy to understand. In the HE state, excitation energy at 488 nm can be used either for emitting a photon or for inducing the cycloreversion reaction back to the LE state. These two processes are mutually exclusive, and as a consequence, the emitted intensity from an ensemble of DAE molecules (which are embedded in a single PS bead) will decrease monotonically as a function of the 488 nm illumination time. The longer the bead is irradiated at 488 nm, the more DAE molecules are converted to the LE state and the lower is the registered fluorescence intensity. As shown in Fig.S4f, even over ten illumination cycles the LE → HE → LE photoconversion reactions are reversible and the cycles can be repeated many times without degradation, testifying that the photochromic reactions of the DAE molecules incorporated into PS are still active.

### 6. Estimation of the number of DAE molecules in a single polymer bead.

To estimate the number of photoswitches that were incorporated into a single polymer bead, we first determined the absorption of a $2.8 \cdot \mu M$ solution of DAE in toluene in both the HE and the LE states, Fig.S5a,b.

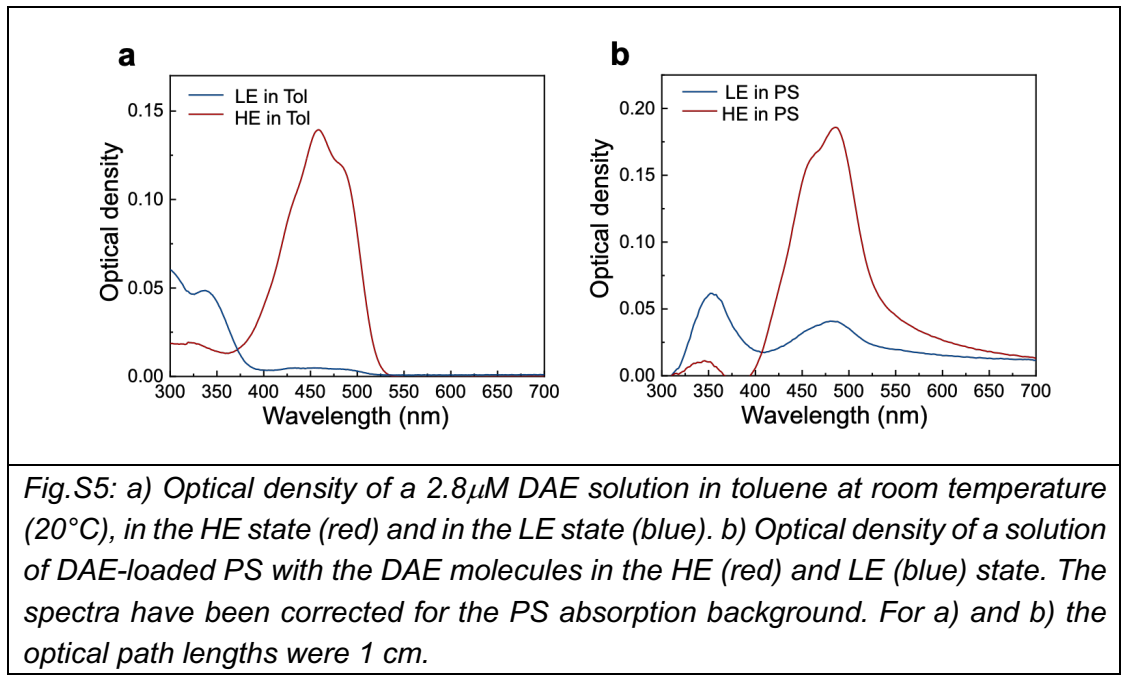

*Fig.S5: a) Optical density of a 2.8μM DAE solution in toluene at room temperature (20°C), in the HE state (red) and in the LE state (blue). b) Optical density of a solution of DAE-loaded PS with the DAE molecules in the HE (red) and LE (blue) state. The spectra have been corrected for the PS absorption background. For a) and b) the optical path lengths were 1 cm.*

The absorption of DAE in the HE state has a small contribution from DAE molecules in the LE state and needs to be corrected using

$$\Delta A_{DTE}^{460,HE} = A_{DTE}^{460,HE,LE} - A_{DTE}^{460,LE},$$

where $A_{DTE}^{460,HE,LE}$ refers to the absorption of DAE at 460 nm in the HE state including the LE contribution (red line fig.S5a), and where $A_{DTE}^{460,LE}$ refers to the absorption of DAE at 460 nm in the LE state (blue line fig.S5a). Using the concentration given above and a cuvette pathlength of 1 cm, this is equivalent to a corrected extinction at 460 nm of

$$\Delta\varepsilon_{DTE}^{460,HE} = \frac{\Delta A_{DTE}^{460,HE}}{c_{DTE} \cdot l} = \frac{A_{DTE}^{460,HE,LE} - A_{DTE}^{460,LE}}{c_{DTE} \cdot l} = \frac{0.13916 - 0.00476}{2.8 \cdot 10^{-6} M \cdot 1 \, cm}$$

$$\approx 48000 \, M^{-1} \cdot cm^{-1}$$

For DAE embedded in PS beads, fig.S4b, a similar procedure yields

$$\Delta A_{DTE \, in \, PS}^{460,HE} = A_{DTE \, in \, PS}^{460,HE,LE} - A_{DTE \, in \, PS}^{460,LE} = 0.145 - 0.0318 = 0.113$$

for the optical density, from which we deduce a DAE concentration of

$$c_{DTE \, in \, PS} = \frac{\Delta A_{DET \, in \, PS}^{460,HE}}{\Delta\varepsilon_{DTE}^{460,HE} \cdot l} = \frac{0.113}{48000 \, M^{-1} \cdot cm^{-1} \cdot 1 \, cm} \approx 2.35 \cdot 10^{-6} \, M$$

This solution was 10 times diluted with respect to the stock solution of DAE in PS. Hence, the DAE concentration in the stock solution of DAE-loaded PS beads amounts to $c_{DTE \, in \, PS_{stock}} = 10 \cdot c_{DTE \, in \, PS} = 2.35 \cdot 10^{-5} \, M$. The stock solution of PS beads was prepared with a molecular weight of $c_{beads_{stock}}^{wt} = 3.3 \cdot 10^{-4} \, kg \cdot L^{-1}$ which corresponds to a molar concentration of $c_{beads_{stock}} = \frac{c_{beads_{stock}}^{wt}}{m_{bead} \cdot N_A}$, where $N_A$ denotes the Avogadro constant. The mass of a bead $m_{bead}$ is obtained from the density $\rho_{bead}$ and the Volume $V_{bead} = \frac{4}{3}\pi r^3$ of the beads. Using $\rho_{bead} = 1000 \, kg \cdot m^{-3}$ and r = 160 nm as provided from the manufacturer, this yields $c_{beads_{stock}} \approx 3.20 \cdot 10^{-11} \, M$. The number of DAE molecules incorporated into a single PS bead, $N_{DTE/bead}$ follows from

$$N_{DTE/bead} = \frac{c_{DTE \, in \, PS_{stock}}}{c_{beads_{stock}}} = \frac{2.35 \cdot 10^{-5} \, M}{3.20 \cdot 10^{-11} \, M} \approx 7.4 \cdot 10^5,$$

corresponding to a molar concentration of

$$c_{DTE/bead} = \frac{N_{DTE/bead}}{N_A \cdot V_{bead}} = \frac{7.4 \cdot 10^5}{6.022 \cdot 10^{23} \, mol^{-1} \cdot 1.72 \cdot 10^{-20} \, m^3} \approx 0.07 \, M$$

DAE molecules within a single bead. Accordingly, the separation of two DAE molecules within a single PS bead amounts on average to about 2.8 nm. For such a small intermolecular distances, quenching effects are expected and we verified this by preparing a series of PS beads with DAE concentrations ranging from 1.2 mM to 71.4 mM. The corresponding spectra and integrated intensities as a function of DAE concentration are shown in fig.6a,b, respectively.

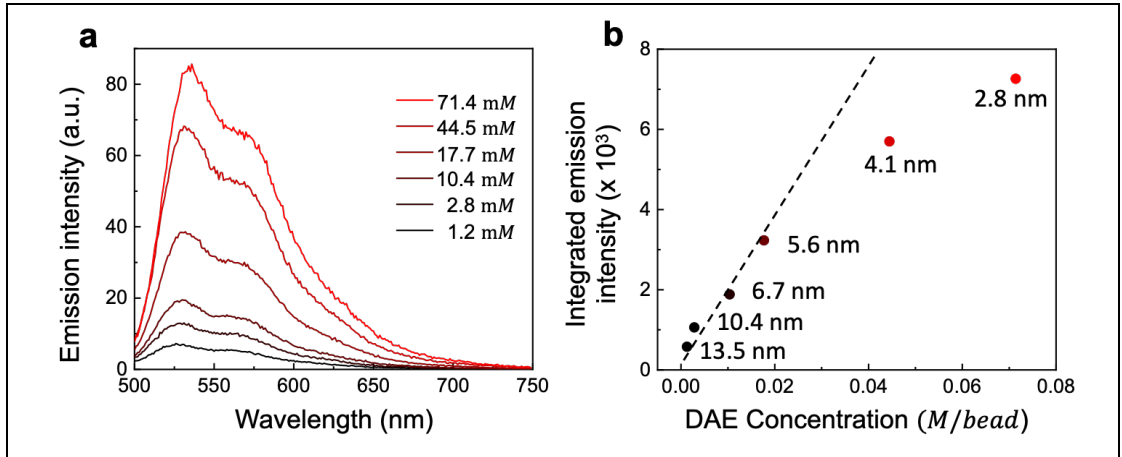

*Fig.S6: a) Emission spectra of DAE-loaded PS beads in the HE state for DAE concentrations per bead from 1.2 mM to 71.4 mM. The excitation wavelength was 488 nm. b) Integrated intensities of the emission spectra shown in a) (same colour code as in a). The dashed line corresponds to a linear relation as a guide for the eye. The numbers next to the data points refer to the average separation of the DAE molecules in the PS beads. The DAE concentration per bead has been obtained from the calculation detailed in the running text.*

The deviation from the linear relationship between the integrated intensity and the DAE concentration per bead in Fig.S6b is obvious. Nevertheless, the highest intensity, which is the parameter we are aiming at, is observed for the highest DAE/bead concentration.

## 7. Microstructuring

The surface templated electrophoretic deposition (STEPD) assembly process of DAE-loaded PS beads followed the same protocol as detailed before [3–5]. Briefly, a 50 nm layer of polymethyl methacrylate (PMMA) resist was spin coated onto an ITO-Glass substrate (Zhuhai Kaivo Optoelectronic Technology Co., Ltd. KV-ITO-P008-2, $\Omega < 7\,\mathrm{ohm/sq}$). Cavities with dimensions of 500 nm × 500 nm separated by 5 μm from each other were prepared in a quadratic grid by electron beam lithography. The patterned template was cut to a piece of 1 cm × 1 cm and mounted in a home-built STEPD cell, that fixes the electrode-to-electrode distance to 2 mm. This results in a sample chamber 5 mm × 5 mm × 2 mm in size for the colloidal solution. The cell is covered with a bare ITO plate that serves as counter electrode. Both electrodes were cleaned by sonication in isopropanol for 1 minute, rinsed with isopropanol and Milli-Q water, and blow-dried with clean nitrogen gas. Subsequently, the cell was sealed with teflon tape to avoid leakages. For assembly of the array of PS beads, 0.5 ml of the solution of DAE-loaded PS beads (0.3 mg/ml) was mixed with a 0.5 ml aqueous solution of NaCl (0.4 mmol/l) and injected into the cell. For the STEPD process the cell was connected to a power supply (Hewlett Packard, E3610A), and a constant DC voltage of 4 V was applied for 5 s. Finally, the surface of the template was gently washed with Milli-Q water and then blow-dried with clean, nitrogen gas.